\documentclass{article}[12pt]
\usepackage[pdftex]{graphicx}   
\begin{document}
\title{Improving approximate vacuum prepared by the adiabatic quantum computation}
\author{Kazuto Oshima\thanks{E-mail: kooshima@gunma-ct.ac.jp}    \\ \\
\sl National Institute of Technology, Gunma College,Maebashi 371-8530, Japan}
\date{}
\maketitle
\begin{abstract}
According to the quantum adiabatic theorem, we can in principle obtain a true vacuum of a quantum system starting from a trivial vacuum of 
a simple Hamiltonian.   In actual adiabatic digital quantum simulation with finite time length and non-infinitesimal time steps, we can
only obtain an approximate vacuum that is supposed to be a superposition of a true vacuum and excited states.  We propose a procedure
to improve the approximate vacuum. 
\end{abstract} 
\\
\newpage
\section{Introduction}
Recently, quantum computers have been developed for studying several systems\cite{Martinez}.
The D-wave is the quantum annealing\cite{Nishimori} machine suitable for solving optimization problems\cite{Farhi}. 
It will be one of the most important subjects for quantum computers to find the lowest energy or the lowest energy
state of a quantum system.    Some quantum systems have been analyzed by digital quantum simulators of universal type. 
The (1+1)-dimensional Schwinger model\cite{Schwinger} is one of the most famous quantum field theory that has been
analyzed by digital quantum simulators\cite{Honda,Okuda}.   One of the bench mark of the analysis is to evaluate
vacuum expectation values of physical quantities.   It has been observed that certain vacuum expectation value
varies in time under a constant Hamiltonian\cite{Honda}.   This means that the approximate vacuum prepared
by the adiabatic quantum computation is not a true vacuum but a superposition of a true vacuum and exited states.
Time average of the vacuum expectation value slightly differs from the exact value that is known for the case the
number of qubits is small\cite{Honda,Oshima}.   Therefore, it is desirable to reduce the affects of the exited states.

The purpose of this paper is to propose a procedure to improve an approximate vacuum reducing affects of
excited states.   For a one-qubit system based on the (1+1)-dimensional Schwinger model, we show a concreate
digital quantum simulation result.    We improve an approximate vacuum and a vacuum expectation value of 
a physical quantity does not varies in time.   We also propose a procedure to improve an approximate vacuum
for the two-qubits case.    Our procedure can be applied successively and approximate precision can in principle be 
improved gradually.    Our procedure can be generalized to an $n$-qubits system.

\section{One-qubit system}
We consider a one-qubit system described by the simple Hamiltonian 
\begin{equation}
{\hat H}=-JH, \quad  J>0,
\end{equation}
where $H$ is the Hadamard gate $H={1 \over 2}(Z+X)$.
This Hamiltonian corresponds to the (1+1)-dimensional Schwinger model with a special
value of a coupling constant formulated on a spatial lattice composed of only two sites.   We construct an approximate vacuum of ${\hat H}$
starting from the trivial vacuum of the initial Hamiltonian ${\hat H}=-JZ$ by the adiabatic quantum computation.    We use the following
slowly time varying Hamiltonian
\begin{equation}
{\hat H}(s)=(1-s){\hat H}_{0}+s{\hat H}, \quad 0 \le s \le 1,
\end{equation}
where $s={t \over T}$ and $T$ is a sufficiently long time length for the adiabatic quantum computation.    For $0 \le t \le T$ we use ${\hat H}(s)$
and for $t \ge T$ we fix ${\hat H}(s)$ as ${\hat H}=-JH$.    According to the adiabatic theorem, we can obtain the true vacuum of ${\hat H}$ at $t=T$,
if $T$ is sufficiently large, ${\hat H}(s)$ changes very slowly with respect to $s$, and the vacuum is not degenerate for ${\hat H(s)}, 0 \le s \le 1$.  
Fig.1 shows a time variation of the expectation value of $Z$ by the
instantaneous vacuum.     In the time region $t \ge T$, a periodic oscillation is observed.  
To the worse, the center of the oscillation
slightly differs from ${1 \over \sqrt{2}}$ that is the exact vacuum expectation value of $Z$.   Therefore, we cannot obtain
the exact value  ${1 \over \sqrt{2}}$ by the time average over a period.    The cause of the oscillation is that the approximate
vacuum $|\psi\rangle$ is a superposition of the true vacuum $|E_{0} \rangle$ and the exited state $|E_{1} \rangle$,
\begin{equation}
|\psi\rangle=\alpha|E_{0} \rangle+\beta|E_{1} \rangle, \quad |\alpha|^{2} +|\beta|^{2}=1,
\end{equation}
where $H|E_{0}\rangle=1$ and $H|E_{1}\rangle=-1$.    We propose a procedure to reduce the value $|\beta|$ and improve the
approximate vacuum.    We prepare one ancilla bit, and put the state $|0\rangle|\psi\rangle$ into the quantum circuit
in Fig.2.   By this quantum circuit the quantum state $|0\rangle|\psi\rangle$ transforms as
\begin{equation}
|0\rangle|\psi\rangle \rightarrow \alpha|0 \rangle|E_{0}\rangle+\beta|1 \rangle|E_{1}\rangle.
\end{equation}
We measure the first qubit by the basis $\{|0 \rangle, |1\rangle \}$.   If we obtain $|0\rangle$, which occurs with the probability $|\alpha|^{2}$,
we measure $Z$ for the second qubit.     If we obtain $|1\rangle$, which occurs with the small probability $|\beta|^{2}$,
we discard the state.    Repeating this measurement many times we can obtain $\langle E_{0}|Z|E_{0}\rangle$ with a sufficient precision.
Fig.3 shows our simulation result, where at the time $t=T$, the quantum circuit in Fig.2 is inserted. 
In Fig.3 we do not adopt the discard procedure for the case $|1\rangle$.   We simulate the value $|\alpha|^{2}\langle E_{0}|Z|E_{0}\rangle
+|\beta|^{2}\langle E_{1}|Z|E_{1}\rangle$, which in our result is 0.706760.   The expectation value $\langle E_{0}|Z|E_{0}\rangle$ can be obtained
by multiplying this value by ${1 \over |\alpha|^{2}-|\beta|^{2}}$ in this case.    Our result is $2|\alpha|^{2}-1=0.999242$ and we have
$0.76760\times{1 \over 0.999242}=0.707296$, which is our approximation value to the exact value ${1 \over \sqrt{2}}$. 

\section{Two-qubits system and generalization}
We propose a procedure to improve an approximate vacuum obtained by the adiabatic quantum computation for a two-qubit system.
We consider the following approximate vacuum
\begin{equation}
|\psi\rangle=\alpha|E_{0}\rangle+\beta|E_{1}\rangle+\gamma|E_{2}\rangle+\delta|E_{3}\rangle, 
\end{equation}
where $|E_{0}\rangle$ is the true vacuum and $|E_{i}\rangle, i=1,2,3$ is the $i$-th exited state.    Our ideal goal is to perform the
following transformation
\begin{equation}
|00\rangle|\psi\rangle \rightarrow \alpha|00\rangle|E_{0}\rangle+\beta|01\rangle|E_{1}\rangle+\gamma|10\rangle|E_{2}\rangle+\delta|11\rangle|E_{3}\rangle.
\end{equation}
After this transformation, measuring the first two qubits, if we obtain $|00\rangle$ we have the true vacuum $|E_{0}\rangle$.   For the system Hamiltonian
${\hat H}$, we set ${\hat U}(\theta)=ie^{-i{ \theta \over 2}{\hat H}}$.  We put $|00\rangle|\psi\rangle$ into the quantum circuit in Fig.4.   We will have the following
transformation
\begin{eqnarray}
|00\rangle|\psi\rangle \rightarrow {1 \over 4}|00\rangle((1+ie^{-i{1 \over 2}E_{0}\theta} +(ie^{-i{1 \over 2}E_{0}\theta})^{2}+(ie^{-i{1 \over 2}E_{0}\theta})^{3})\alpha|E_{0}\rangle \nonumber \\
+(1+ie^{-i{1 \over 2}E_{1}\theta} +(ie^{-i{1 \over 2}E_{1}\theta})^{2}+(ie^{-i{1 \over 2}E_{1}\theta})^{3})\beta|E_{1}\rangle \nonumber \\
+(1+ie^{-i{1 \over 2}E_{2}\theta} +(ie^{-i{1 \over 2}E_{2}\theta})^{2}+(ie^{-i{1 \over 2}E_{2}\theta})^{3})\beta|E_{2}\rangle \nonumber \\
+(1+ie^{-i{1 \over 2}E_{3}\theta} +(ie^{-i{1 \over 2}E_{3}\theta})^{2}+(ie^{-i{1 \over 2}E_{3}\theta})^{3})\beta|E_{3}\rangle) \nonumber \\
+ {1 \over 4}|01\rangle(\cdots)+ {1 \over 4}|10\rangle(\cdots)+ {1 \over 4}|11\rangle(\cdots).
\end{eqnarray}
We define an approximation value of $E_{0}$ by $E_{0}^{\prime}=\langle \psi|{\hat H}|\psi\rangle$, which can be estimated by a simple quantum simulation.
We set the parameter $\theta$ by $E_{0}^{\prime}\theta=\pi$.    Since $E_{0}^{\prime}$ is close to $E_{0}$  the coefficient of $\alpha|00\rangle|E_{0}\rangle$
in Eq.(7) will satisfy  ${1 \over 4}|1+ie^{-i{1 \over 2}E_{0}\theta} +(ie^{-i{1 \over 2}E_{0}\theta})^{2}+(ie^{-i{1 \over 2}E_{0}\theta})^{3}| \approx 1$, and the coefficients
of $\beta|00\rangle|E_{1}\rangle, \gamma|00\rangle|E_{2}\rangle$ and $\delta|00\rangle|E_{3}\rangle$ are expected to be small compared with 1.
We measure the first two qubits.    The state $|00\rangle$ will be obtained with high probability.   We discard other states $|01 \rangle, |10\rangle$ and $|11\rangle$.
Thus we have obtained an approximate vacuum more close to the true vacuum $|E_{0}\rangle$.    This procedure can in principle be repeated ad infinitum with improving
the value $E_{0}^{\prime}$.

For an $n$-qubits system, we prepare $n$ ancilla qubits in the state $|0\rangle|0\rangle\cdots|0\rangle$.    Using the Hadamard gate on the ancilla qubits
and acting the controlled-${\hat U}(\theta)$ gates between each ancilla bit and an $n$-qubits approximate vacuum(Fig.5), we can improve the approximate
vacuum obtained by the adiabatic quantum computation.

\section{Summary}
We have proposed a procedure to improve an approximate vacuum obtained by the adiabatic quantum computation.
We have shown a concreate quantum simulation result for a simple Hamiltonian in one-qubit system.
We also have given a quantum circuit to improve an approximate vacuum for a two-qubits system.
Our procedure can easily be generalaized to an $n$-qubits system.

\newpage
Figure Captions\\
\\
Fig.1 \quad Vacuum expectation value of $Z$ by qasm-simulator of IBM.   We set $J={\pi \over 4}, T=36$ and one time-step width is ${1 \over 24}$.
We use $10^{6}$ shots.\\
\\ \\
Fig.2 (a) \quad Quantum circuit that separates the true vacuum and the exited state. \qquad (b) \quad Concreate quantum circuit of (a) consists of fundamental
gates in Qiskit; $Rz(\theta)=e^{-i{\theta \over 2}Z},  Rx(\theta)=e^{-i{\theta \over 2}X}$ and  $S=|0\rangle\langle 0|+i|1\rangle\langle 1|$.\\
\\ \\
Fig.3 \quad   Vacuum expectation value of $Z$.    At the time $t=T=36$, the quantum circuit in Fig.2 is inserted.    The main cause of
the discontinuity at $t=T$ derives mainly from the vanishment of the cross term ${\rm Re}({\bar \alpha}\beta\langle E_{0}|Z|E_{1}\rangle)$.\\
\\ \\
Fig.4 \quad Quantum circuit that approximately separates the true vacuum and the exited states.\\
\\ \\ 
Fig.5 \quad Quantum circuit that approximately separates the true vacuum and the exited states for an $n$-qubits system.\\
\\
\newpage
{} \qquad \\ \\ \\ \\
{\quad }  \includegraphics[width=2cm]{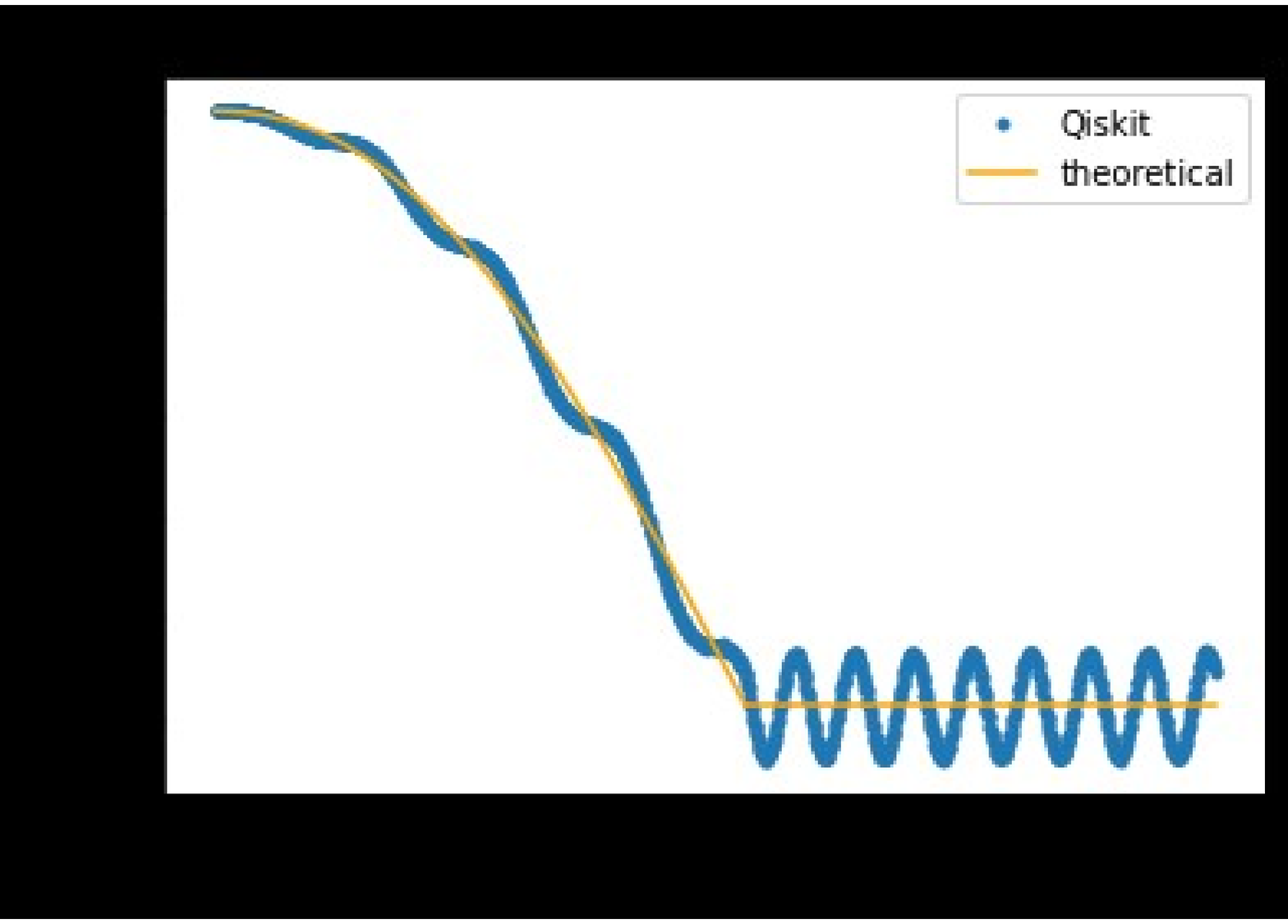}
Fig.1\\
\vspace{10pt}
\\
\includegraphics[width=6cm]{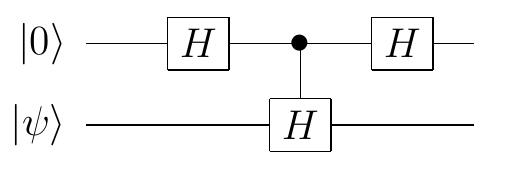} \qquad 
\includegraphics[width=8cm]{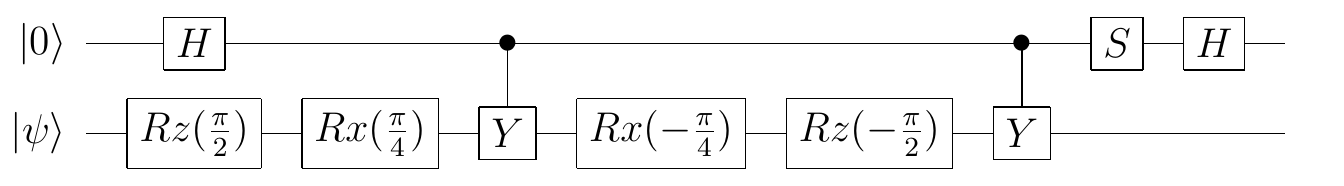}   \\
\\
Fig. 2 \qquad \qquad \qquad  (a) \qquad \qquad \qquad \qquad \qquad \qquad \qquad \qquad  \qquad  (b)\\
\\
\vspace{0pt}
{\quad }  \includegraphics[width=8cm]{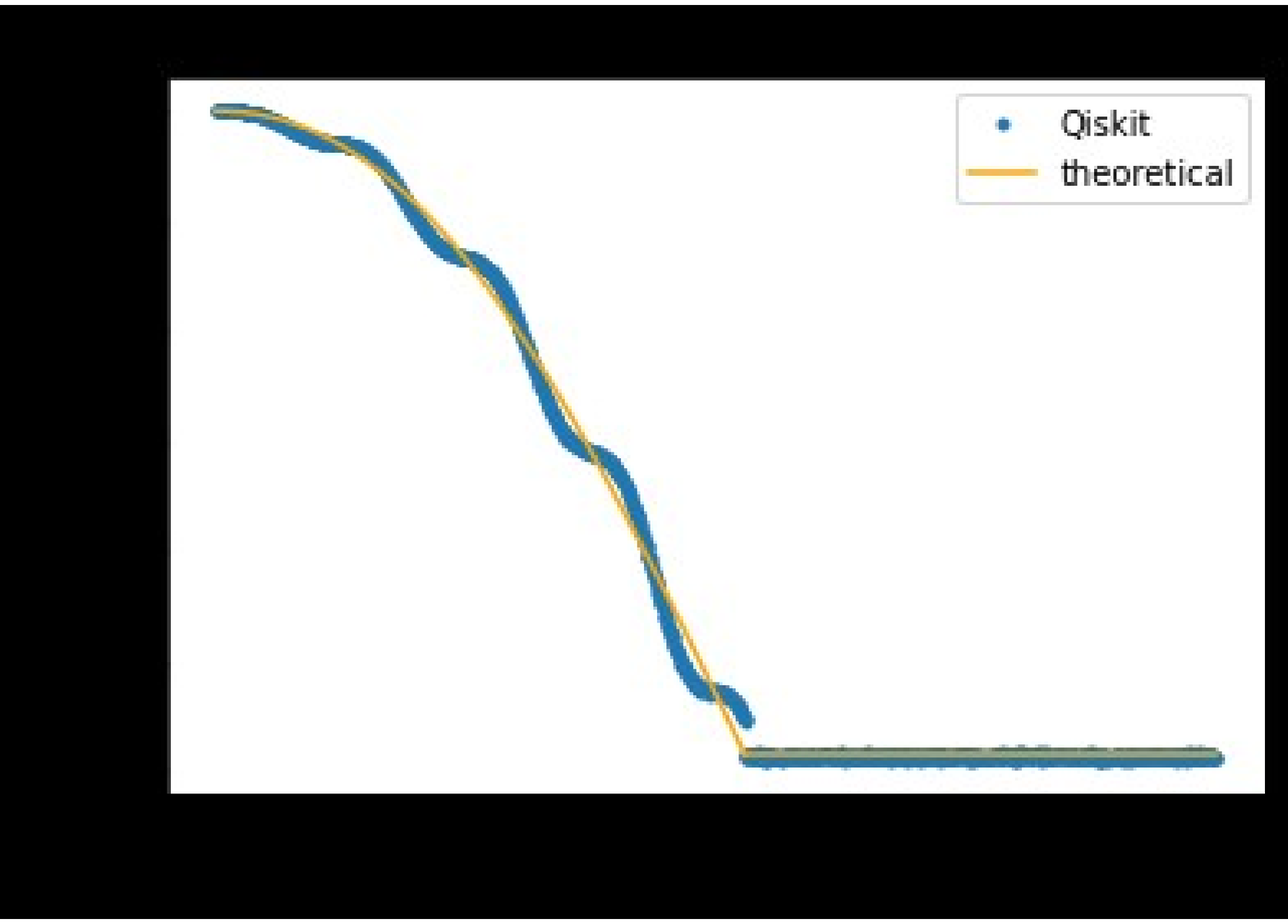}
Fig.3\\
\\
\includegraphics[width=8cm]{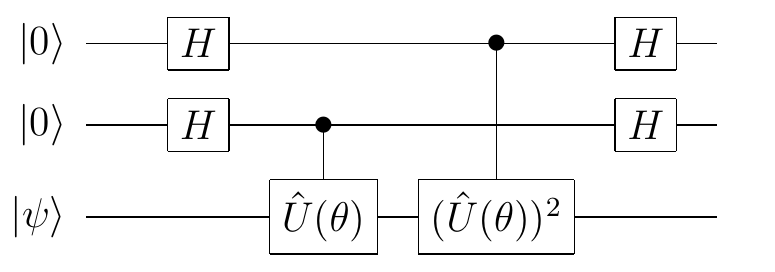}  \qquad 
\includegraphics[width=8cm]{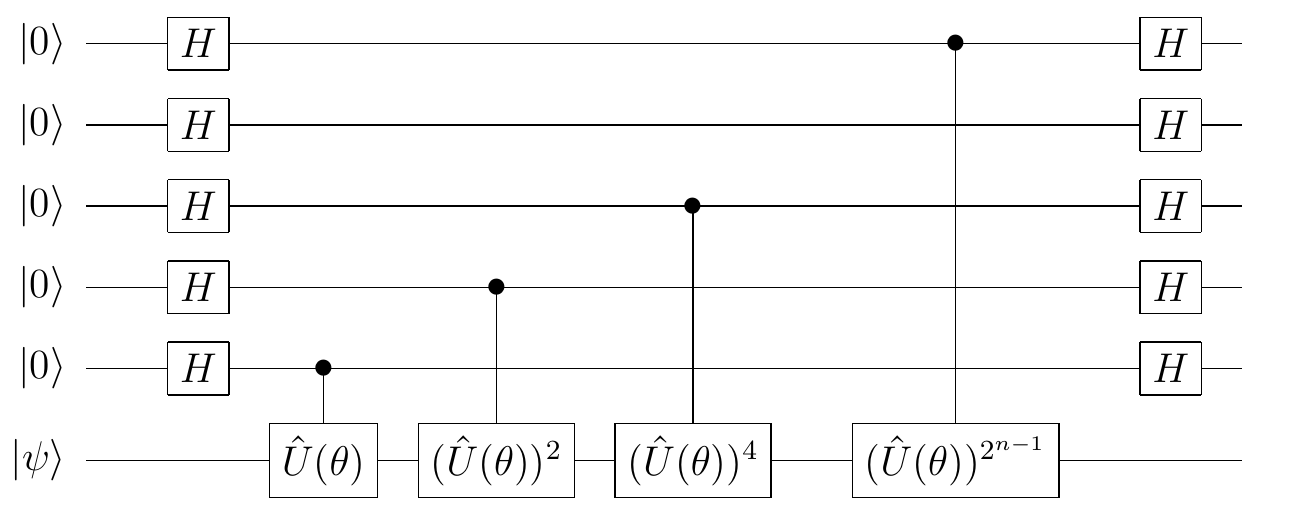} 
\\ 
{} \qquad \qquad \qquad \qquad \qquad  Fig.4 \qquad \qquad \qquad \qquad \qquad \qquad \qquad \qquad \qquad \qquad  \qquad  Fig.5\\

\end{document}